\documentclass{aa}  

\usepackage{graphicx}
\usepackage{txfonts}
\usepackage{subscript}
\usepackage{makecell}
\begin{document}

   \title{Simple molecules and complex chemistry in a protoplanetary disk}

   \subtitle{A JWST investigation of the highly inclined disk d216-0939}
   
   \titlerunning{Ice Chemistry in d216-0939 with JWST}

   \author{A.~Potapov
          \inst{1}\fnmsep\thanks{Corresponding author}
          \and
          H.~Linz\inst{1,2}
          \and
          J.~Bouwman
          \inst{2}
          \and
          W.~Rocha
          \inst{3}
          \and
          J.~Martin
          \inst{4}
          \and
          S.~Wolf
          \inst{4}
          \and
          Th.~Henning
          \inst{2}
          \and
          H.~Terada
          \inst{5}
          }

   \institute{Analytical Mineralogy Group, Institute of Geosciences, 
   Friedrich Schiller University Jena, Jena, Germany \\
   \email{alexey.potapov@uni-jena.de}
         \and
   Max Planck Institute for Astronomy, Heidelberg, Germany\\
   \email{[linz,bouwman]@mpia.de}
         \and
   Laboratory for Astrophysics, Leiden Observatory, Leiden University, 
   Leiden, The Netherlands
   %\email{rocha@strw.leidenuniv.nl}
         \and
   Institute for Theoretical Physics and Astrophysics, Christian Albrecht 
   University Kiel, Kiel, Germany 
   %\email{[jmartin,wolf]@astrophysik.uni-kiel.de}
         \and
   National Astronomical Observatory of Japan, Tokyo, Japan 
   %\email{terada@naoj.org}
             }

   \date{Received December 10, 2024; accepted February 20, 2025}

\abstract{While the number of detected molecules, particularly complex organic molecules, in the solid-state in astrophysical environments is still rather limited, laboratory experiments and astrochemical models predict many potential candidates. Detection of molecules in protoplanetary disks provides a bridge between the chemical evolution of the interstellar medium and the chemistry of planets and their atmospheres.}
{The excellent spectral sensitivity, broad wavelength coverage and high spatial resolution of the James Webb Space Telescope (JWST) allows for making progress in exploring chemical compositions of various astrophysical environments including planet-forming disks. They are a prerequisite for probing the disk content by means of sensitive absorption studies.} 
{In this paper, we present initial results of the JWST Cycle 1 GO program 1741 on d216-0939, a highly inclined TTauri disk located in the outskirts of the Orion Nebula Cluster. We utilise the NIRSpec and MIRI integral field unit spectrographs to cover its spectrum from 1.7 to 28~$\mu$m.} 
{In the d216-0939 disk, we give assignments of the composition of silicate grains. We unambiguously detect solid-state features of H$_2$O,  CO$_2$, $^{13}$CO$_2$, CO, OCN$^-$, and tentatively OCS; species that had been detected recently also in other circumstellar disks. For the first time in disks, we provide unique detections of ices carrying NH$_4^+$ and the complex organic molecule ammonium carbamate (NH$_4^+$NH$_2$COO$^-$).}
{The latter detections speak for a very efficient NH$_3$ chemistry in the disk. We also show the very important role of scattering in the analysis of observational spectra of highly inclined disks.}


   \maketitle

\section{Introduction}

In cold dense astrophysical environments, such as molecular clouds, protostellar envelopes and planet-forming disks beyond the snowline, the solid state is presented by dust grains mixed with molecular ices. The full composition of interstellar and circumstellar ices is still an open question. Water is the main constituent accounting for more than 60\% of the ice in most lines of sight \citep[e.g., ][]{Whittet2003,Dishoeck2021}. In addition to water, several other molecules have been unambiguously detected in ices: CO, OCS,  CO$_2$, NH$_3$, H$_2$CO, HCOOH, CH$_4$, and CH$_3$OH \citep{McGuire2022} and recently OCN$^-$, NH$_4^+$, and tentatively SO$_2$ \citep{McClure2023}. Basing on results of experimental and modelling studies simulating conditions and chemistry of astrophysical environments, we know that chemical reactions in/on such ices triggered by UV photons, heat, atoms, and cosmic rays may lead to the formation of many complex organic molecules (COMs) including direct precursors of prebiotic species and prebiotic species themselves, such as sugars, amino acids, and nucleobases. Here, we refer the reader to several review papers \citep{Theule2013,Linnartz2015,Oberg2016, Arumainayagam2019,Sandford2020,Potapov2021a,Fulvio2021}. \\
However, detection of COMs in interstellar and circumstellar ices (in the solid state) is still a challenge because these molecules are less abundant, and their IR features are less distinct. In interstellar ice, methanol (CH$_3$OH), has been unambiguously detected \citep{Pontoppidan2003}. In the pre-JWST era, tentative detections of acetaldehyde \citep[CH$_3$CHO][]{Schutte1999}, ethanol \citep[C$_2$H$_5$OH][]{Terwisscha2018}, and urea \citep[H$_2$NCONH$_2$][]{Raunier2004} have been proposed. New JWST data provide first tentative \citep[e.g., ][]{Yang2022} and then statistically robust \citep{Rocha2024} detections of C$_2$H$_5$OH, CH$_3$CHO and methyl formate (HCOOCH$_3$), as well as tentatively acetonitrile \citep[CH$_3$CN,][]{McClure2023} and ethyl cyanide \citep[C$_2$H$_5$CN,][]{Nazari2024} in interstellar ices.

Detection of molecules in planet-forming disks provides a bridge between the chemical evolution of the interstellar medium (ISM) and chemistries of planets and their atmospheres. Recent JWST/NIRSpec observations of the edge-on Class {\sc ii} protoplanetary disk HH 48 NE revealed spatially resolved absorption features of the major ice components H$_2$O,  CO$_2$, and CO, multiple weaker signatures from less abundant ices NH$_3$, OCN$^-$, and OCS and, for the first time in a protoplanetary disk, isotopologue $^{13}$CO$_2$ ice \citep{Sturm2023a}. JWST/MIRI observations of the same disk provided evidence for the presence of CH$_4$ and CH$_3$OH \citep{Sturm2023b}.

Despite these recent successes regarding smaller species, the general number of detected ice species in disks, particularly COMs, is still very limited. However, according to the laboratory experiments and astrochemical models there are many potential candidates. This manuscript  presents results of the initial analysis of JWST  observations of the protoplanetary disk d216-0939, a highly inclined disk in the outskirts of the Orion Nebula Cluster, and provides assignments of the ice and dust bands detected.

\section{Observations, data reduction and analysis}

   \begin{figure*}
    \hspace*{-0.5cm}
      \begin{minipage}[b]{0.60\textwidth}
    \includegraphics[width=\textwidth]{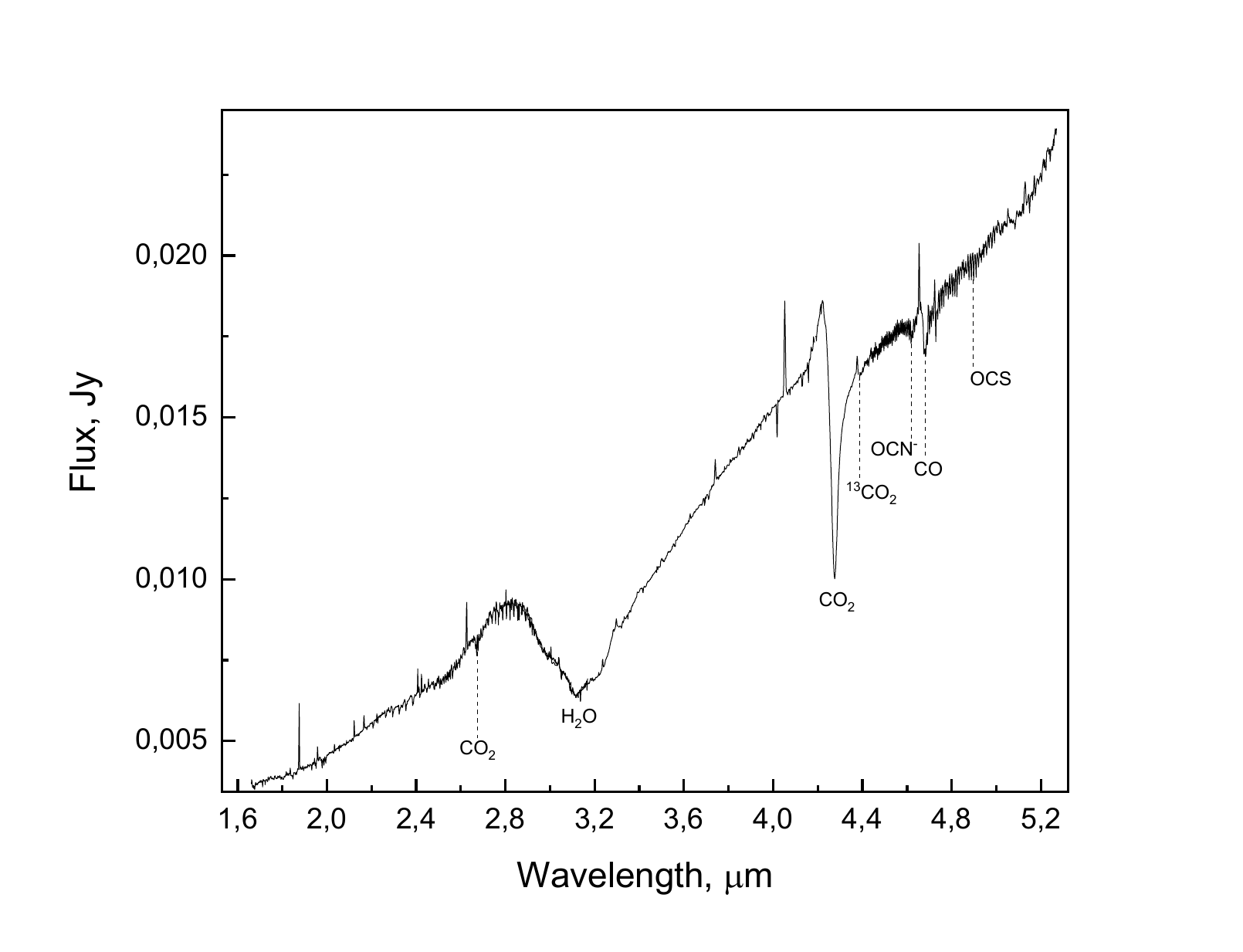}
  \end{minipage}
  \hspace*{-2cm}
  \begin{minipage}[t]{0.60\textwidth}
    \includegraphics[width=\textwidth]{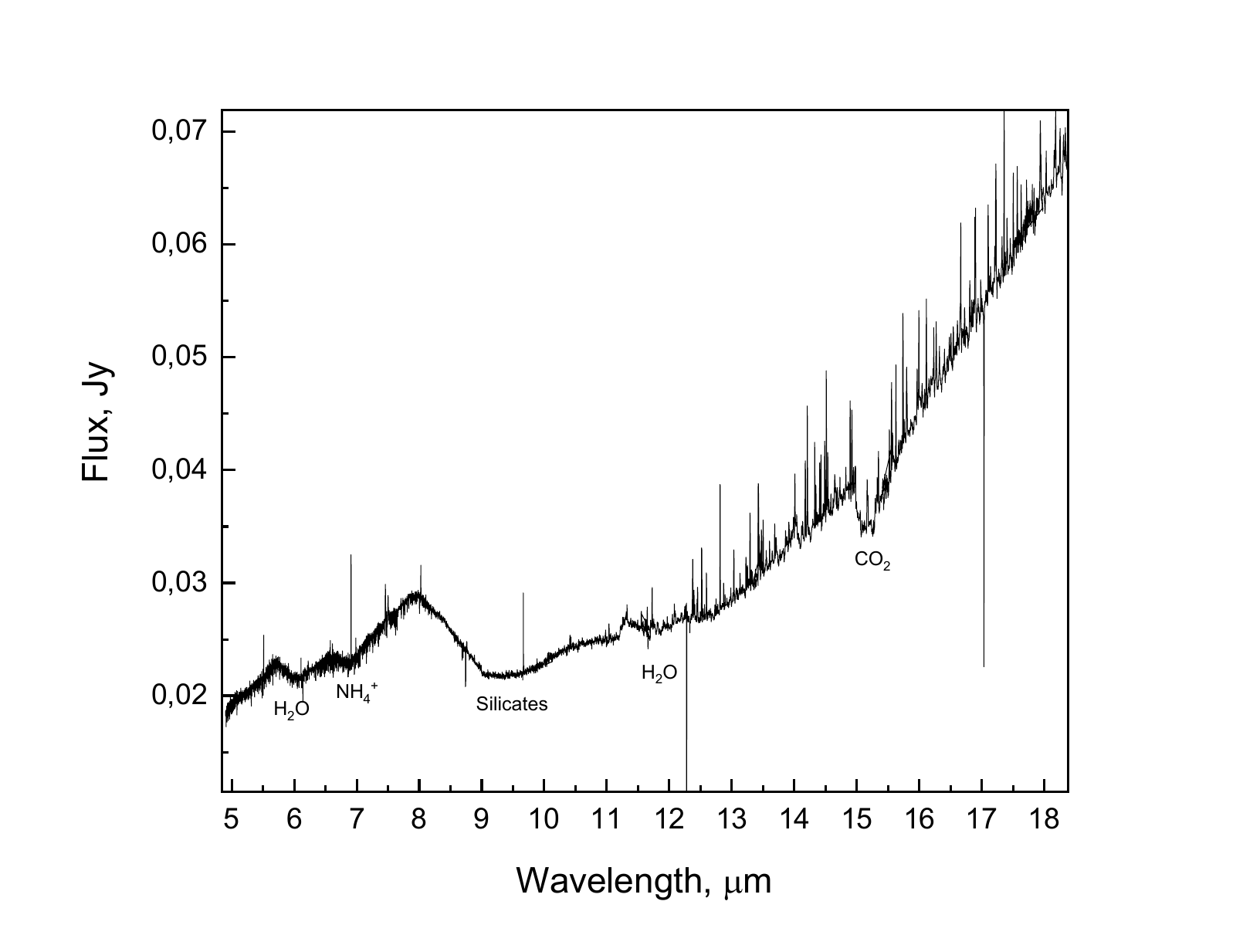}
  \end{minipage}
      \caption{NIRSpec (left) + MIRI (right) spectrum of the d216-0939 disk and assignments of the main spectral bands. Gas phase lines are not labelled.
              }
         \label{Fig:Full-spectrum}
   \end{figure*}

\subsection{Source selection}

Ice bands have been observed in emission in planet-forming disks at far infrared wavelengths \citep[e.g., ][]{Min2016}. However, to observe the near- and mid-infrared ice bands in emission, one would typically need temperatures of several hundred Kelvin, which is far above the sublimation temperature. The best way to observe ice bands at near-infrared wavelengths is through absorption line studies of highly inclined disks \citep[e.g., ][]{Terada2012a}. At high inclination angles the light of the central star is attenuated by the circumstellar disks allowing us to probe a wide region of the disk. \\
For our study, we selected the highly inclined disk d216-0939 (a.k.a. V$^*$ V2377 Ori or \object{HOPS 65}) in Orion, a high-mass star forming region likely to resemble the birthplace of the Solar System \citep[cf.][]{Young2020}. In the Hubble Space Telescope (HST) paper by \citet{Smith2005} this disk system stood out as featuring a large silhouette disk in HST optical imaging. A high-resolution ALMA study has derived an inclination of the outer disk of around 75--78$^\circ$ \citep{Sheehan2020}. The selected target is among the few known suitably inclined disks that show indications of crystalline water ice \citep{Terada2012a,Terada2012b} and strong reddening of the central star by the circumstellar disk material. As crystalline water ice forms at temperatures reached close to the snowline, this selection criterion means that we are likely probing a sightline in this system across the snowline. However, other scenarios may exist. Crystallization may also occur earlier in the history of the system when the outer disk was warmer, or material may have been transported outwards after being crystallized near the snowline. We note furthermore that the locations of snowlines are not constant over the evolution of the disk due to a variety of circumstances \cite[e.g., ][]{Martin2012,Owen2020,Mori2021,Ros2024}.

\subsection{JWST observations}

The highly inclined protoplanetary disk in Orion d216-0939 was observed with the integral field units (IFUs) of both the Near-Infrared Spectrograph  \citep[NIRSpec, ][]{Jakobsen2022,Boeker2022} and the Mid-Infrared Instrument \citep[MIRI, ][]{Wright2023} within our General Observation (GO) programme 1741 (PI: A.~Potapov) at the end of the JWST Cycle 1. The NIRSpec observations used the medium resolution settings G235M/F170LP and G395M/F290LP to cover the 2 to 5-micron wavelength range, and especially to seamlessly cover the full extent of the water ice feature between 2.7 and 3.5 micron. As the target is expected to be slightly extended at the shortest wavelengths of NIRSpec we used a 4-point dither. Given the IFU size we apply a target acquisition (TA) for the NIRSpec observations on 2 nearby stars to make sure our targets are properly centred on the image slicer. We used nearby stars with good-quality GAIA information for the TA as the target itself could show some deviation from point-source geometry at short wavelengths. As the IFU size of MIRI is large enough, no TA was needed. Observations with the MIRI Medium Resolution Spectrograph (MRS) were done for all 3 grating settings so the entire wavelength range between 5 and 28 microns was continuously covered. In the MIRI range we were especially interested in the silicate absorption features at 10 and 18 microns, as well as several faint absorption signals of complex organic molecule ices, mainly in the 5 - 10-micron range. For the MIRI observations we applied a 4-point dither, to get a proper spatial and wavelength sampling. At the longer wavelengths the source is expected to be compact, meaning the background can be estimated from the observations itself and no additional off-pointing was required.

We opted for medium resolution spectroscopy for both instruments, since the offered spectral resolution in the low-res mode (R\textasciitilde100) is too coarse to allow a robust spectral decomposition of the H$_2$O ice features at the shortest wavelengths, and, in case of MIRI do not cover all required wavelengths
(\textgreater{} 10 microns). This also facilitated a proper detection and separation of gas-phase lines (e.g., from water) from the ice features. We employed the IFUs of both instruments which provides a better control over the spatial distribution of the (continuum) emission/absorption and ice feature absorption signals.

\subsection{Data reduction}\label{Section:Data reduction}

All data reduction was performed with the JWST pipeline \citep[see][]{Bushouse2024} version 1.13.4, and was based on the pmap1210 calibration scheme. We processed the raw data with slightly modified pipeline scripts. Spectral de-fringing was applied to all data sets. We relied on the extract1D step to extract the 1D spectrum from the 2D spectral cubes. This extracts aperture photometry per cube slice. An aperture correction is applied afterwards. Both the aperture and the aperture correction are wavelength-dependent. For MIRI we experimented with aperture scaling between 1.5 ... 2.0 $\times$ FWHM and found that the smaller aperture does not change the continuum shape, but gives slightly higher signal-to-noise. For NIRSpec, the aperture size scaling within extract1D is fixed (to 2 $\times$ FWHM). For most of the data, we could use the autocenter function to enable a precise centroiding of the aperture on source before spectral extraction. We noticed that at the shortest wavelengths covered ($\lambda \lesssim \ 2.5 \mu$m) the source more and more deviates from a point-source geometry, probably showing scattering at the more extended disk surface. We checked that the deviation from the point-source nature is negligible for the longer wavelengths where the important spectral features reside. Hence, an extraction approach relying on point source aperture corrections seems still warranted.

\subsection{Approach for fitting the measurements}\label{Section:Fitting Approach}

We do not have a well-constrained photospheric model for the central star, yet. Moreover, the IR continuum may be dominated by thermal emission from warm dust in the inner disk rather than from the stellar photosphere. At this point in the project we refrained from trying to develop such a detailed radiative transfer model for the star-disk system that could guide the determination of the continuum baseline. As a pragmatic approach, we therefore used a third-order polynomial fit for the NIRSpec range with continuum points of 2.5 - 2.55, 2.6 - 2.65, 2.69 - 2.80, 4.0 - 4.04, 4.06 - 4.07, 4.6 - 5.0, 5.15 - 5.2 $\mu$m and two linear fits for the MIRI ranges 5.72 - 7.88 and 8 -- 15.9 $\mu$m. The uncertainties introduced by the continuum do not influence the band assignments presented in the paper. \\

We used the ENIIGMA fitting tool \citep{Rocha2021} to fit multiple features across the NIRSpec and MIRI spectral ranges by scaling {laboratory spectra in optical depth scale of} dust/ice and ice mixtures and pure ices to match the continuum-subtracted optical depths. {This methodology is already described in different papers, such as \citet{Rocha2021, Rocha2024, Rocha2025}. Briefly, these optical depth spectra ($\tau$) are calculated as:}
\begin{equation}
    \tau_{\nu}^{lab} = 2.3 \ Abs_{\nu},
\end{equation}
{where $Abs_{\nu}$ is the laboratory data (absorbance) as function of the wavenumber ($\nu$) in units of cm$^{-1}$. \\
Some of the ices also play a role in the scattering, which is needed to correctly reproduce the observed spectral features. 
We applied the computational code {\sc OpTool} \citep{Dominik2021} to investigate the scattering. While {\sc OpTool} is in principle capable of computing absorption and scattering opacities, we use the tool strictly for deriving the scattering opacities of different silicate grains covered with a variety of ices and ice mixtures. For the absorption opacities, we rely on the laboratory-measured ice data directly.

The relation between optical depth and opacity is given by:}
\begin{equation}
    \tau_{\nu} = a \ \kappa_{\nu},
\end{equation}
{where $a$ is a scaling factor.}

{In {\sc OpTool} }, we utilized the default approach implemented therein, i.e., Distribution of Hollow Spheres (DHS) with $f_{max} = 0.8$, with compact grains (porosity set to zero) and ice mantles of different compositions. We adopted an upper size limit of 1 $\mu$m and a power law of 3.5 for the grain size distribution. This seems to be a realistic approach; however, we note that \citet{Dartois2024} found that the grain sizes can significantly alter the shapes of the ice features. We experimented with the grain size distributions, increasing the upper size limit to 2 and 3 $\mu$m. This did not, however, lead to a fit improvement. Furthermore, it worsened the overall SED fit we attempted.\\

\begin{table*}
\caption{EMIIGMA fitting results for the 2.8 -- 4.0 $\mu$m wavelength region.}             
\label{table:1}      
\centering          
\begin{tabular}{l c c c c c c c c }     
\hline\hline     
 & \multicolumn{8}{c}{Combination $\#$}  \\ 
 Component and Fit & 1 & 2 & 3 & 4 & 5 & 6 & 7 & \textbf{8}  \\
 \hline \\
MgSiO$_3$/H$_2$O 150 K (A)  & x & x & x & x & x & x &   &    \\[1mm]
MgSiO$_3$/H$_2$O 150 K (S)  & x & x & x & x & x & x & x & \textbf{x}  \\[1mm]
NH$_4^+$/OCN$^-$ 80 K       &   & x &   & x &   &   &   &    \\[1mm]
``Chemistry''               &   &   & x & x & x & x & x & \textbf{x}  \\[1mm]
MgSiO$_3$/H$_2$O 10 K (A)   &   &   &   &   & x &   & x &    \\[1mm]
MgSiO$_3$/H$_2$O 10 K (S)   &   &   &   &   & x &   &   &    \\[1mm]
MgSiO$_3$/H$_2$O 100 K (A)  &   &   &   &   &   & x &   & \textbf{x}  \\[1mm]
MgSiO$_3$/H$_2$O 100 K (S)  &   &   &   &   &   & x &   &    \\[1mm]
AIC                         & 19.1& 9.8 & 7.4 & 9.5 & 11.3 & 11.4 & 7.4 & \textbf{7.2} \\[1mm]
RMSE                        & 0.055 & 0.028 & 0.016 & 0.016 & 0.015 & 0.015 & 0.016 & \textbf{0.015} \\
\hline                 
\end{tabular}
\tablefoot{The Akaike Information Criterion (AIC) and Root-mean-square error (RMSE) for various input mixtures are listed. The best fit parameters are marked in boldface.}
\end{table*}

{We used two metrics to evaluate the goodness of the fit, the Root Mean Square Error (RMSE) and the Akaike Information Criterion (AIC). While the RMSE measures the average difference between model and observations, the AIC can quantify the quality of the fit based on the number of parameters used for the fit. The $AIC$ is given by:}   
\begin{equation}
   AIC = \chi^2 + 2p + \frac{2p(p+1)}{N - p - 1}, 
\end{equation}
{where $p$ is the number of free parameters and $N$ is the sample size. $\chi^2$ is given by:}
\begin{equation}
    \chi^2 = \frac{1}{dof}\sum_{i=0}^{n-1} \left(\frac{\tau_{\nu,i}^{\rm{obs}}  - \sum_{j=0}^{m-1} w_j \tau_{\nu,j}^{\rm{lab}}}{\gamma_{\nu,i}^{\rm{obs}}} \right)^2,
\end{equation}
{where $dof$ is the number of degrees of freedom (i.e. $N - p$), $m$ is the number of laboratory spectra used in the fit, $w_j$ the scaling factor applied to the individual spectra, $\gamma$ is the error in the observational optical depth spectrum. Once $AIC$ is calculated, different models can be compared using the $\Delta$ AIC, which is defined as $AIC_i$ - $min(AIC)$, where $AIC_i$ represents values for different models, and $min(AIC)$ corresponds to the minimum AIC value. $\Delta$AIC below 2 means a substantial support of the model, while $\Delta$ AIC clearly above 2 indicates considerably less or no support of the model \citep{burnham2002model}. In this paper, the lowest RMSE and AIC is considered the best model, which is not necessarily the model containing more components.}

\section{Results}

In Figure~\ref{Fig:Full-spectrum}, we present the complete NIRSpec + MIRI spectrum up to 18 $\mu$m of the d216-0939 disk.

In the following, we split the spectrum into regions and discuss our assignments of the dust and ice bands done using the {\sc ENIIGMA} fitting tool, from short to long wavelengths starting at the 3-micron region containing the stretching vibration band of H$_2$O and ending at the 8 -- 16-micron region containing spectral signatures of silicates. We performed separate fits because determining a single global continuum for disks is rather challenging without an accurate radiative transfer model.

\subsection{2.8 -- 4.0-micron region}

The main spectral band in this region is the H$_2$O stretching mode. It may also contain signatures of NH$_3$, CH$_3$OH, NH$_4^+$, and complex organic molecules. Particularly, ammonium salts including ammonium carbamate (NH$_4^+$NH$_2$COO$^-$) and ammonium formate (HCOO$^-$NH$_4^+$) may explain the red part of the band as was shown by the analysis of the Rosetta spectra of the comet 67P \citep{Poch2020}.

Based on our previous results demonstrating dust/ice mixing in cold astrophysical environments \cite{Potapov2021b}, we  use spectral data obtained for physical mixtures of silicates and water ice. Crystalline water ice was previously detected in the d216-0939 disk \citep{Terada2012a,Potapov2021b} and its presence was one of the main parameters while choosing the target for our JWST observations.

It was also clear from the analysis of the  CO$_2$ stretching band at 4.27 $\mu$m that scattering plays a crucial role in the short-wavelength region (see Sect.~\ref{Section:4.0-4.5micron}). The scattering component  of the H$_2$O band was obtained utilising {\sc OpTool}, with the optical constants as inputs obtained for the MgSiO$_3$/H$_2$O mixtures with the mass ratio of 2.7 and presented in \citet{Potapov2018b}. The absorption spectra of these mixtures were also used as absorption components of the fits. 

\begin{figure}
   \hspace*{-0.5cm}
   \includegraphics[width=4.1in]{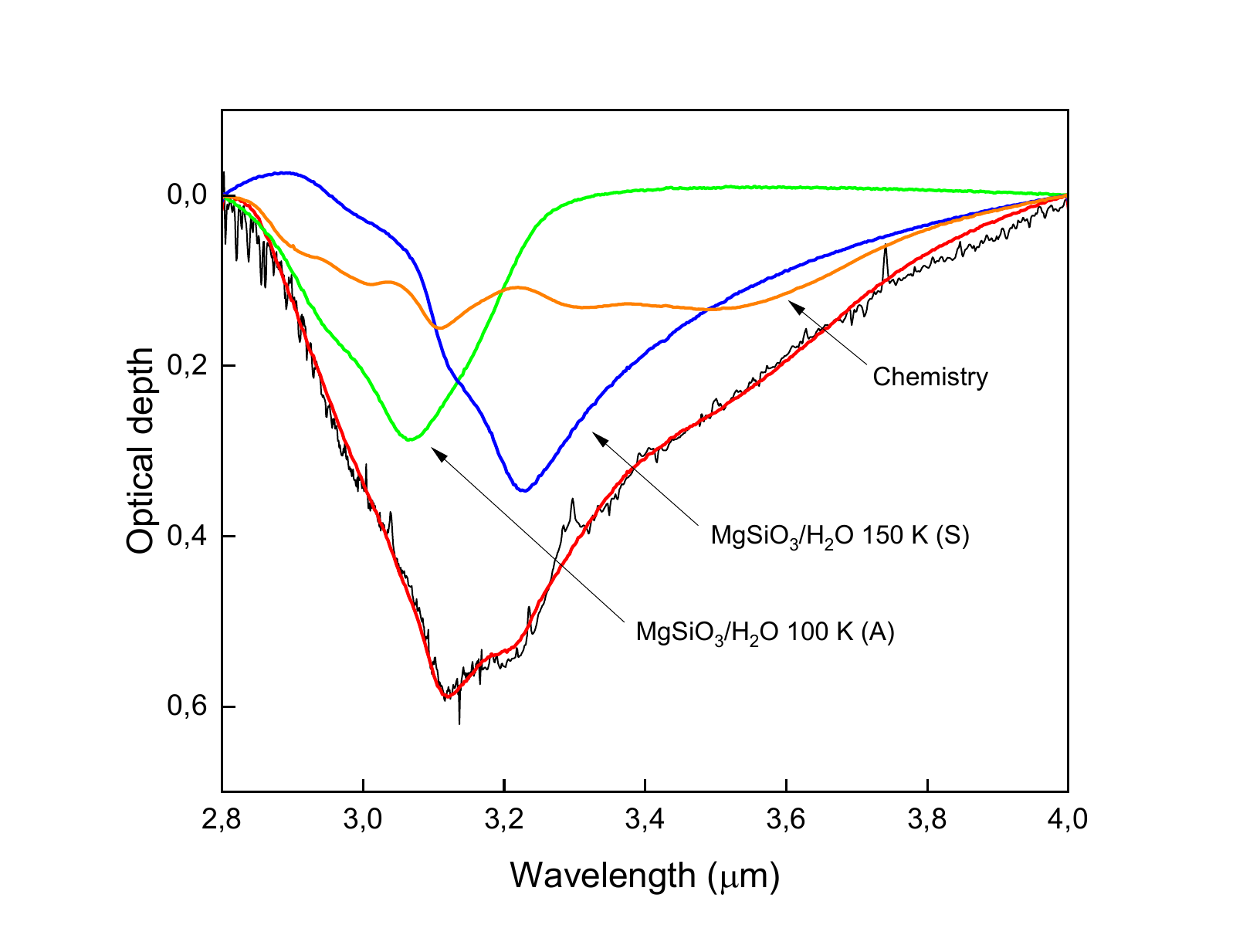}
      \caption{Observational spectrum (black), the best fit model for the 2.8 -- 4.0 $\mu$m region (red) and its components:
MgSiO$_3$/H$_2$O 150 K (S) (blue),
MgSiO$_3$/H$_2$O 100 K (A) (green), and
``chemistry'' (orange).
              }
         \label{Fig:2.8-4.0}
   \end{figure}

The best fit model is shown in Figure~\ref{Fig:2.8-4.0} and was obtained using the spectra of MgSiO$_3$/H$_2$O 150 K scattering (S) + MgSiO$_3$/H$_2$O 100 K absorption (A) + ``chemistry''. ``Chemistry'' is the laboratory spectrum obtained after UV irradiation of a mixture of  CO$_2$ + NH$_3$ at 75 K and its subsequent heating to 230 K \citep{Potapov2022}. The spectrum mainly contains spectral signatures of ammonium carbamate and carbamic acid (NH$_2$COOH). The high temperature (230 K) spectrum was chosen to reduce the contribution of NH$_3$, which has strong bands at 3 and 9 $\mu$m but is not unambiguously detected.

The Akaike Information Criterion (AIC) and Root-mean-square error (RMSE) were used to estimate the quality of the fits and are presented in Table \ref{table:1}. The criteria for the best fit were: (i) decrease of the RMSE and (ii) increase of the AIC by not more than 2 with respect to the lowest value. We started with the 150 K absorption (A) and scattering (S) spectra as ENIIGMA inputs. We do not see clear spectral signatures of NH$_3$ and CH$_3$OH in the 10-micron range, where the species have the strongest bands. NH$_4^+$ is detected at 6.85 $\mu$m (see below), and addition of the NH$_4^+$/OCN$^-$ 80 K spectrum \citep[originally presented in][]{Novozamsky2001}  from the LIDA database \citep{Rocha2022} improved the fit considerably. However, addition of the ``chemistry'' spectrum to the fit provided a much better fit and was necessary to fit the red shoulder of the band. The peaks at 3.1 and 3.3 $\mu$m coincide with the peaks of several ammonium salts (ammonium formate, ammonium citrate, ammonium sulphate, ammonium carbamate), see \citet{Poch2020}. In our case, ammonium carbamate was detected in the ``chemistry'' sample \citep{Potapov2022}. The 3.3 $\mu$m peak is clearly affected by PAH emission. Addition of the NH$_4^+$/OCN$^-$ spectrum to the fit after ``chemistry'' led to an increase of the AIC by more than 2 as a consequence of the ignorance of the NH$_4^+$/OCN$^-$ spectrum by the fit.

At the next step, we added MgSiO$_3$/H$_2$O 10 and 100 K (A) and (S) inputs to consider the temperature gradient in the disk. Increase of the AIC in both cases indicated that not all spectra were considered by the fit. The ignored spectra were in both cases low-temperature scattering and high-temperature absorption. Finally, the best fit was obtained for the MgSiO$_3$/H$_2$O 150 K (S) + MgSiO$_3$/H$_2$O 100 K (A) + ``chemistry'' inputs. We note that the usage of MgSiO$_3$/H$_2$O 150 K (A) or MgSiO$_3$/H$_2$O 10 K (A) instead of MgSiO$_3$/H$_2$O 100 K (A) also provides a reasonable fit. Thus, 150 K scattering provides a defining contribution to the spectrum.

\subsection{4.0 -- 4.5-micron region}\label{Section:4.0-4.5micron}

\begin{table*}
\caption{Input data sets for fitting the  CO$_2$ region around 4.27 $\mu$m.}             
\label{table:2}      
\centering          
\begin{tabular}{l c c }     
\hline\hline 
Detailed Description & Rocha Model & Trivial name for Table~\ref{table:3}  \\   
\hline                    
H$_2$O: CO$_2$:NH$_3$:CH$_4$ 10:1:1:1 72 K absorbance & D13a & Warm absorption \\
H$_2$O: CO$_2$:NH$_3$:CH$_4$ 10:1:1:1 35 K absorbance & D12a & Lukewarm absorption \\
H$_2$O: CO$_2$ 10:1 13 K absorbance & D2a & Cold absorption \\
CO$_2$ 13 K absorbance & G2 & Cold absorption pure CO$_2$ \\
MgSiO$_3$/H$_2$O 100K core + D13a 72 K mantle scatter & D13a & Warm scattering \\
MgSiO$_3$/H$_2$O 10K core + D12a 35 K mantle scatter & D12a & Lukewarm scattering \\
MgSiO$_3$/H$_2$O 10K core + D2a 13 K mantle scatter & D2a & Cold scattering \\
MgSiO$_3$/H$_2$O 10K core + G2 13 K mantle scatter & G2 & Cold scattering pure  CO$_2$ \\
\hline                  
\end{tabular}
\tablefoot{The paper references are \citet{Rocha2014} for model G2, and \citet{Rocha2017} for models D13a, D12a, and D2a.}
\end{table*}

\begin{table*}
\caption{EMIIGMA fitting results for the 4.0 -- 4.5 $\mu$m wavelength region.}             
\label{table:3}        
\begin{tabular}{l c c c c c c c c c c c c}   
\hline\hline     
 & \multicolumn{12}{c}{Combination $\#$}  \\ 
 Component and Fit                    & 1 & 2 & 3 & 4 & 5 & \textbf{6} & 7 & 8 & 9 & 10 & 11 & 12  \\
 \hline \\
Warm absorption                       & x &   &   &   & x &            & x &   & x & x  &    &    \\[1mm]
Lukewarm absorption                   &   & x &   &   &   & \textbf{x} &   & x &   &    &    &    \\[1mm]
Cold absorption                       &   &   & x & x &   &            &   &   & x & x  & x  & x  \\[1mm]
Cold absorption pure CO$_2$           &   &   &   &   &   &            &   &   &   & x  & x  & (x)\\[1mm]
Warm scattering                       & x &   &   &   &   &            & x &   & x & x  &    &    \\[1mm]
Lukewarm scattering                   &   & x &   &   &   &            &   & x &   &    &    &    \\[1mm]
Cold scattering                       &   &   & x & x & x & \textbf{x} & x & x & x & x  & x  & x  \\[1mm]
Cold scattering pure CO$_2$           &   &   &   & x & x & \textbf{x} & x & x & x & x  & x  & x  \\[1mm]
AIC & 8.09& 9.60 & 6.55 & 7.05 & 7.10 & \textbf{7.02} & 9.02 & 9.13 & 11.06 & 14.22 &  9.08  & 9.17 \\[1mm]
RMSE & 0.029 & 0.033 & 0.026 & 0.014 & 0.015 & \textbf{0.014} & 0.014 & 0.015 & 0.014 & 0.020  & 0.014 & 0.015 \\
\hline                
\end{tabular}
\tablefoot{The Akaike Information Criterion (AIC) and Root-mean-square error (RMSE) for various input mixtures. The best fit parameters are marked in boldface.}
\end{table*}

This spectral region contains the strong absorption feature of $^{12}$CO$_2$ ice at 4.27 $\mu$m and the weaker ice feature of the $^{13}$CO$_2$ variety at 4.39 $\mu$m. Note that we also detected the combination mode of $^{12}$CO$_2$ at 2.68 $\mu$m. Very interesting is the positive bump blueward of the $^{12}$CO$_2$ absorption feature at 4.27 $\mu$m. This bump is relatively broad, though it does not coincide in wavelength with known solid-state feature of ice, dust or PAHs. We assume that here we see the imprint of scattering on ice mantles covering dust grains. This effect has been described and modelled by \citet{Dartois2022}.

\begin{figure}
   \centering
   \includegraphics[width=3.5in]{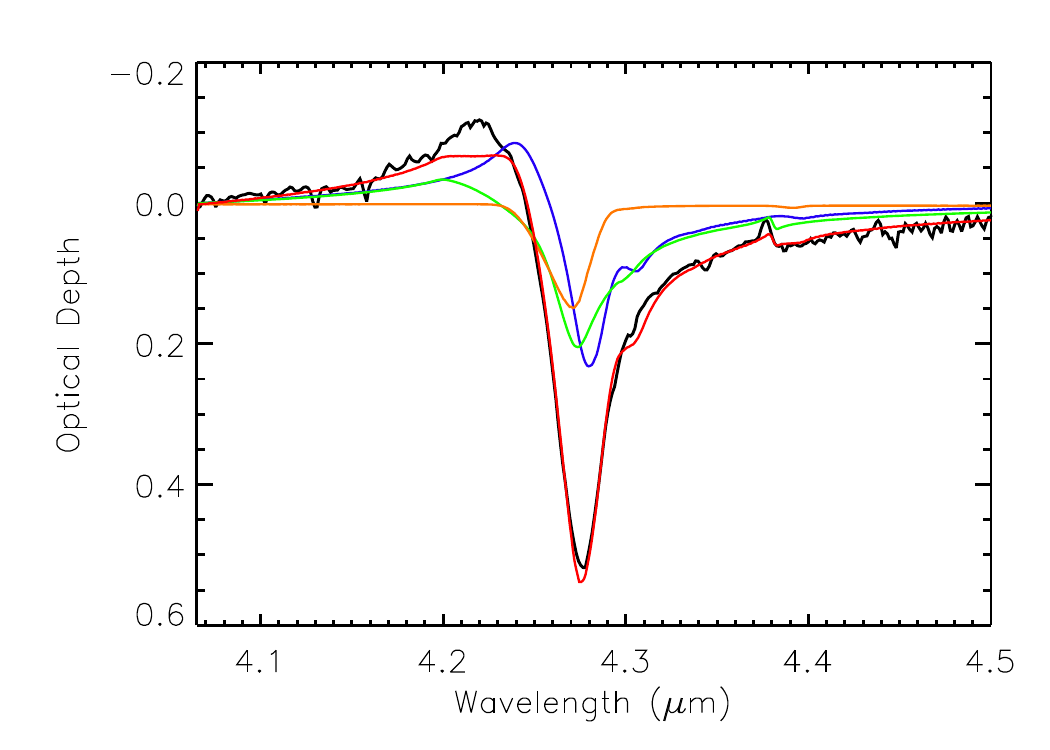}
      \caption{Observations (black), the best fit model for the 4.0 -- 4.5 $\mu$m region (red) and its components: blue -- cold scattering on 13 K H$_2$O: CO$_2$ mantles, green -- cold scattering on 13 K pure  CO$_2$ mantles, orange -- absorption in 35 K H$_2$O: CO$_2$ mantles.
              }
         \label{Fig:4.0-4.5}
   \end{figure}

The best fit model is shown in Figure~\ref{Fig:4.0-4.5} and was obtained using the spectra of scattering on 13 K H$_2$O: CO$_2$ mantles, cold scattering on 13 K pure  CO$_2$ mantles, and
absorption in 35 K H$_2$O: CO$_2$ mantles.

For modelling the main  CO$_2$ feature, we started with pure absorption data. A detailed overview of the utilised absorbance and optical constants is given in Table~\ref{table:2}. We first consider moderate-temperature ice mixtures where  CO$_2$ is mixed with H$_2$O. Absorbance data exist for temperatures of 13 K (cold), 35 K (lukewarm), and 72 K (warm). It quickly turns out that such pure absorbance reference data cannot replicate the full shape of the measured  CO$_2$ feature well. They are still too narrow, they cannot capture the wide red wing, and obviously do not contain a positive (``negative absorption'') feature that could emulate the bump close to 4.2 $\mu$m. We therefore decided to include extinction caused by scattering also for the  CO$_2$ fitting into the selection of input data for {\sc ENIIGMA}. We used {\sc OpTool} to compute the scattering properties with the model parameters as described in Section~\ref{Section:Fitting Approach}. To be consistent, we used optical constants for the grain core and mantle at identical or at least similar temperatures, respectively. For the grain core, we used the same MgSiO$_3$/H$_2$O data as above. For the mantle material, we used the optical constants from several databases (cf.~Table~\ref{table:2}).  We excluded the wavelength range containing the Brackett~$\alpha$ line at around 4.05 $\mu$m and start at 4.065 $\mu$m.

We continued with two-component fits of absorbance + scattering at similar temperatures, respectively. The red absorption wing is much better fitted by the inclusion of scattering. Additionally, the scattering produces some positive feature blueward of the main absorption feature, although the exact shape does not agree. The fit continuously improves when going from warm absorption + warm scattering to cold absorption + cold scattering. Another clear step of improvement occurs when including scattering of a pure cold  CO$_2$ mantle. Such data offer an additional positive component that helps to fit better the truly blue part of the positive bump. Furthermore, the faint $^{13}$CO$_2$ absorption feature\footnote{We note in passing that also the $^{13}$CO$_2$ feature at around 4.39~$\mu$m seems to show a positive (scattering) bump blueward of its absorption maximum, similar to the one discussed for the $^{12}$CO$_2$ feature.} at 4.39 $\mu$m is better captured when the pure CO$_2$ data are included. (The CO$_2$ samples used for the laboratory reference data contain the $^{13}$C and $^{12}$C isotopologues in a ratio of 1:100\,.)  It must be noted that the pure  CO$_2$ component was also required to fit the 10-micron range (see Sect.~\ref{Sect:8-16}).

When more than three components are combined, several acceptable solutions with roughly similar fit quality and RMSE are possible, however, AIC is increased by more than 2 in all the cases. One notes that in many cases, the warm and lukewarm scattering components are ignored by {\sc ENIIGMA}. Inclusion of one variety of cold scattering is, however, mandatory to reach a good fit. Our modelling is not fully conclusive regarding the necessity to have (luke)warm absorption and (luke)warm scattering. If given a multitude of input choices, {\sc ENIIGMA} will include a 72 K absorption component in the final fit, possibly since those warm spectra have less of a wiggle at around 4.3 $\mu$m (probably an $^{18}$OCO feature) than cold absorption models from the used spectral libraries. The AIC and RMSE values are presented in Table~\ref{table:3}.

\subsection{4.5 --4.8-micron region}

\begin{table*}
\caption{EMIIGMA fitting results for the 4.5 -- 4.8 $\mu$m wavelength region.}             
\label{table:4}      
\centering          
\begin{tabular}{l c c c c c c c c c c}   
\hline\hline     
 & \multicolumn{10}{c}{Combination $\#$}  \\ 
 Component and Fit                  & 1 & 2 & 3 & 4 & 5 & 6 & 7 & 8 & 9 & \textbf{10}  \\
 \hline \\
NH$_4^+$/OCN$^-$ 12 K               & x &   &   &   &   &   &   &   &   &   \\[1mm]
pure CO 15 K                        & x & x &   &   &   & x &   & x &   &  \textbf{(x)}  \\[1mm]
NH$_4^+$/OCN$^-$ 80 K               &   & x & x & x & x & x & x & x & x &  \textbf{x}   \\[1mm]
\makecell[l]{H$_2$O:CO$_2$:CO:
NH$_3$:CH$_3$OH 15K}                &   &   & x &   & x & x &   &   &   &    \\[1mm]
C/H$_2$O                            &   &   &   & x & x & x &   &   &   &    \\[1mm]
Rocha D8a H$_2$O:NH$_3$:CO 13 K     &   &   &   &   &   &   & x & x & x &  \textbf{x}  \\[1mm]
cold Rocha D8a-mantle scattering    &   &   &   &   &   &   &   &   & x &  \textbf{x}  \\[1mm]
AIC                         & 7.44& 7.39 & 6.89 & 7.55 & 8.91 & 10.81 & 6.79 & 8.81 & 7.20 & \textbf{9.18 } \\[1mm]
RMSE                        & 0.027 & 0.026 & 0.024 & 0.027 & 0.024 & 0.024 & 0.024 & 0.024 & 0.0151 & \textbf{0.0148} \\
\hline  \\                
\end{tabular}
\tablefoot{The Akaike Information Criterion (AIC) and Root-mean-square error (RMSE) for various input mixtures are listed. The best fit parameters are marked in boldface.}
\end{table*}

\begin{figure}
   \centering
   \includegraphics[width=3.5in]{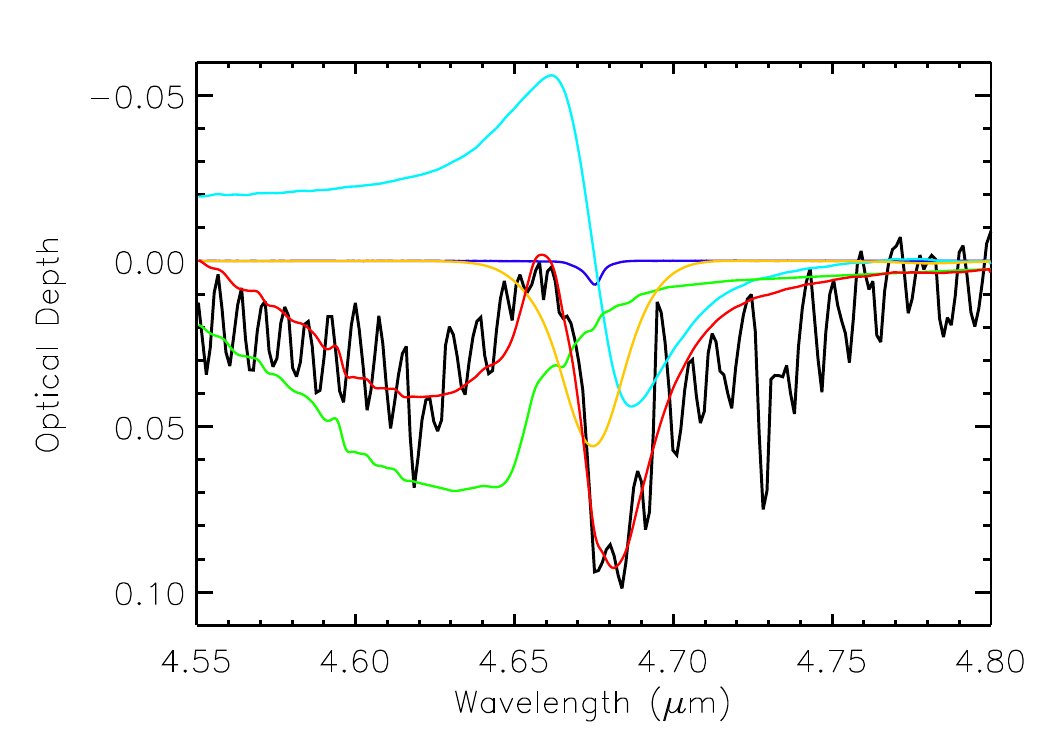}
      \caption{Observations (black), the best fit model for the 4.5 -- 4.8 $\mu$m region (red) and its components: dark blue -- pure CO absorption (15 K), light blue -- scattering on H$_2$O:CO ice mantle (13 K), green -- NH$_4^+$/OCN$^-$ 80 K, yellow -- absorption on H$_2$O:CO ice mantle (13 K).
              }
         \label{Fig:4.5-4.8}
   \end{figure}

This spectral region contains the spectral signatures of OCN$^-$ at 4.62 $\mu$m and CO at 4.67 $\mu$m. We fitted these two features together in ENIIGMA but removed the strong Pfund $\beta$ line at 4.65 $\mu$m from the optical depth input data before fitting, since it turned out to be confusing for the fitting routine. The OCN$^-$ feature is in the NH$_4^+$/OCN$^-$ spectrum from LIDA database. The best fit model is presented in Figure~\ref{Fig:4.5-4.8}. The AIC and RMSE values are presented in Table~\ref{table:4}.

There are data for NH$_4^+$/OCN$^-$ at 12 K and 80 K. We tested both data sets. The cold 12 K data come, however, with a quite pronounced secondary absorption feature at around 4.65 $\mu$m which is not observed in our data. The 80 K file does not show this and is smoother and better fitting to the JWST measurements. Also, a preference for the 80 K spectrum is a clear result of the 5.6 -- 8 $\mu$m region fitting (see subsection~\ref{Sect:5.6-8.0}).

Pure CO absorption data from LIDA give a rather sharp narrow absorption feature at around 4.67 $\mu$m, while our measured CO absorption profile is wider. We therefore also use CO in a H$_2$O mixture. As one example, we use the absorbance of a H$_2$O:CO$_2$:CO:NH$_3$:CH$_3$OH mixture at 15~K taken from LIDA. As a further choice, we utilize the model ``D8a'' lab data from \citet{Rocha2017}, which represents an H$_2$O:NH$_3$:CO mixture with 10:6:4 mass ratio, at 13 K. With such a dataset in the mix, the core of the absorption is filled out better. The general shape, especially the absorption wing at longer wavelengths, however, is not fully replicated. Hence, we included scattering on ice mantles also for the CO range. We used the aforementioned model D8a which has optical constants available to compute the scattering properties, taking the MgSiO$_3$/H$_2$O material properties at 10 K for the grain core. The inclusion of scattering widens the profile and fills the red wing satisfactorily. We note that the pure CO absorption is quite strongly down-weighted in the four-component mix that gives a good fitting result. The inclusion of pure CO seems to improve the RMSE to a small degree. Still, we cannot exclude the possibility that the narrow pure CO absorption just contributes to better fitting some absorption in the deepest absorption trough of the global feature that may arise from superimposed roto-vibrational absorption lines of CO seen in the spectrum.
\noindent
We also tentatively detected OCS at 4.9 $\mu$m.

\subsection{5.6 -- 8-micron region}\label{Sect:5.6-8.0}

\begin{table*}
\caption{EMIIGMA fitting results for the 5.6 -- 8.0 $\mu$m wavelength region.}             
\label{table:5}      
\centering          
\begin{tabular}{l c c c c c c c }    
\hline\hline     
 & \multicolumn{7}{c}{Combination $\#$}  \\ 
 Component and Fit                  & 1 & 2 & 3 & 4 & \textbf{5} & 6 & 7   \\
 \hline \\
MgSiO$_3$/H$_2$O 100 K              & x &   & x &   &            &   &      \\[1mm]
NH$_4^+$ 12 K                       & x & x &   &   &            &   &      \\[1mm]
MgSiO$_3$/H$_2$O 150 K              &   & x &   & x & \textbf{x} & x & x     \\[1mm]
NH$_4^+$ 80 K                       &   &   & x & x & \textbf{x} & x & x    \\[1mm]
``Chemistry''                       &   &   &   &   & \textbf{x} & x & x     \\[1mm]
Hydrocarbons                        &   &   &   &   &            & x &      \\[1mm]
CH$_4$                              &   &   &   &   &            &   & x     \\[1mm]
AIC                                 & 16.2 & 14.5 & 12.7 & 11.4 & \textbf{8.1} & 10.2 & 10.1   \\[1mm]
RMSE                                & 0.049 & 0.046 & 0.042 & 0.038 & \textbf{0.020} & 0.020 & 0.020  \\
\hline  \\                
\end{tabular}
\tablefoot{The Akaike Information Criterion (AIC) and Root-mean-square error (RMSE) for various input mixtures are listed. The best fit parameters are marked in boldface.}
\end{table*}

   \begin{figure}
   \hspace*{-0.5cm}
   \includegraphics[width=4.0in]{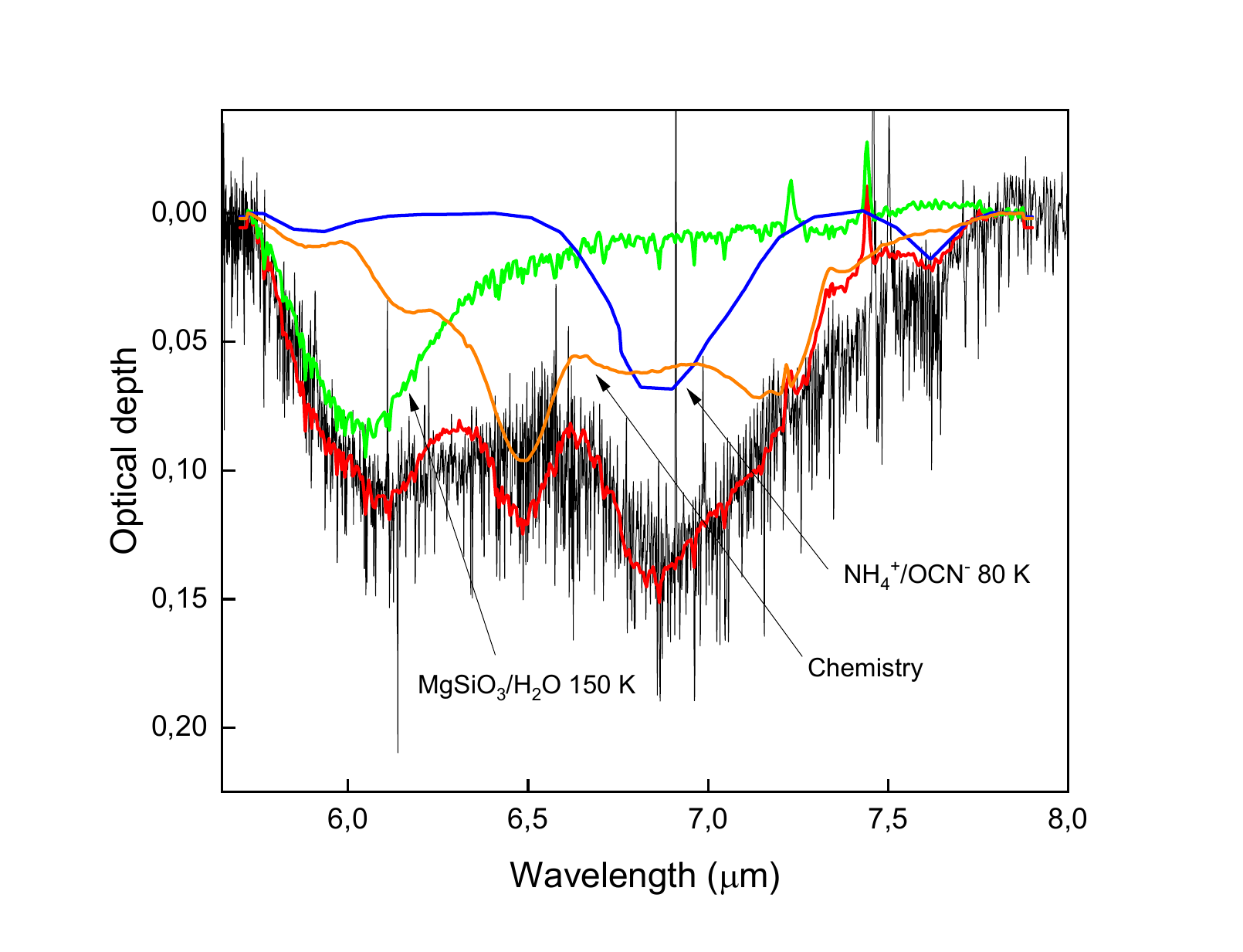}
      \caption{Observations (black), the best fit model for the 5.6 -- 8 $\mu$m region (red) and its components: MgSiO$_3$/H$_2$O 150 K (green) + NH$_4^+$/OCN$^-$ 80 K (blue) + ``chemistry'' (orange).
              }
         \label{Fig:5.6-8.0-1}
   \end{figure}

This spectral region contains the H$_2$O bending vibration mode at 6.2 $\mu$m, the NH$_4^+$ band at 6.9 $\mu$m, and the OCN$^-$ band at 7.6 $\mu$m. The best fit model is presented in Figure~\ref{Fig:5.6-8.0-1} and was obtained using the input mixture of the spectra of MgSiO$_3$/H$_2$O (A) 150 K + NH$_4^+$/OCN$^-$ 80~K + ``chemistry''.

   \begin{figure}
   \hspace*{-0.5cm}
   \includegraphics[width=4.0in]{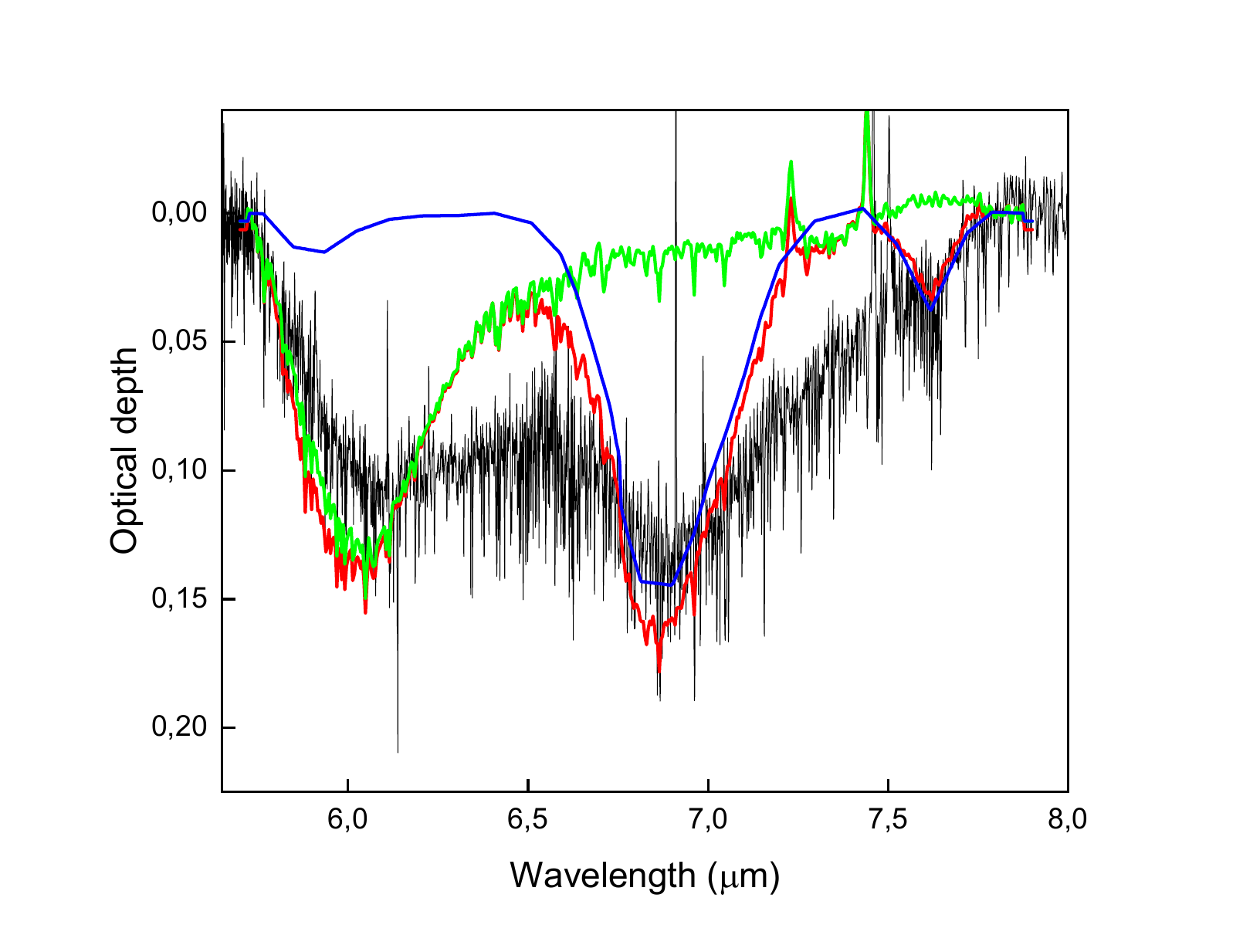}
      \caption{Intermediate model for the 5.6 -- 8 $\mu$m region including MgSiO$_3$/H$_2$O 150 K (green) and NH$_4^+$/OCN$^-$ 80 K (blue).
              }
         \label{Fig:5.6-8.0-2}
   \end{figure}

We started with the MgSiO$_3$/H$_2$O 100 and 150 K and NH$_4^+$/OCN$^-$ 12 and 80 K inputs. There is a red shift of the NH$_4^+$ band with temperature. Our fits clearly show that the 12 K band does not match the observational band, and that the 80 K spectrum must be used. This was the reason for trying the NH$_4^+$/OCN$^-$ 80 K spectrum also in the 3-micron range. The usage of the MgSiO$_3$/H$_2$O 150 K spectrum provided a better fit and was used for further fits. This intermediate result is presented in Fig.~\ref{Fig:5.6-8.0-2} where the observational spectrum is fitted by two laboratory spectra, MgSiO$_3$/H$_2$O 150 K and NH$_4^+$/OCN$^-$ 80 K. The NH$_4^+$ and OCN$^-$ bands (6.85 and 7.6 $\mu$m correspondingly) are fitted well but it is obvious that there is missing signal between 6.2 and 6.7 and between 7.0 and 7.6 $\mu$m.

Exactly in these regions (6.2 -- 6.7 and 7.0 -- 7.3 $\mu$m), ammonium carbamate
(NH$_4^+$NH$_2$COO$^-$), the product of  CO$_2$ + NH$_3$ chemistry, has spectral signatures \citep{Bossa2008,Potapov2019}. The next input to the model was the ``chemistry'' spectrum, which contains the aforementioned bands of ammonium carbamate. This led to the best fit shown in Fig.~\ref{Fig:5.6-8.0-1}. An excess signal around 7.5 $\mu$m in the observational spectrum points to possible additional components. As potential candidates we considered refractory hydrocarbons obtained by UV irradiation of CH$_3$OH ice \citep[the spectrum was presented in][]{Potapov2021b}, as well as CH$_4$ from LIDA database \citep[presented in][] {Rachid2020}. Additions of these components did not improve the fit. The AIC and RMSE values are presented in Table~\ref{table:5}.

\subsection{8 -- 16-micron region}\label{Sect:8-16}

\begin{figure}
   \hspace*{-0.5cm}
   \includegraphics[width=4.0in]{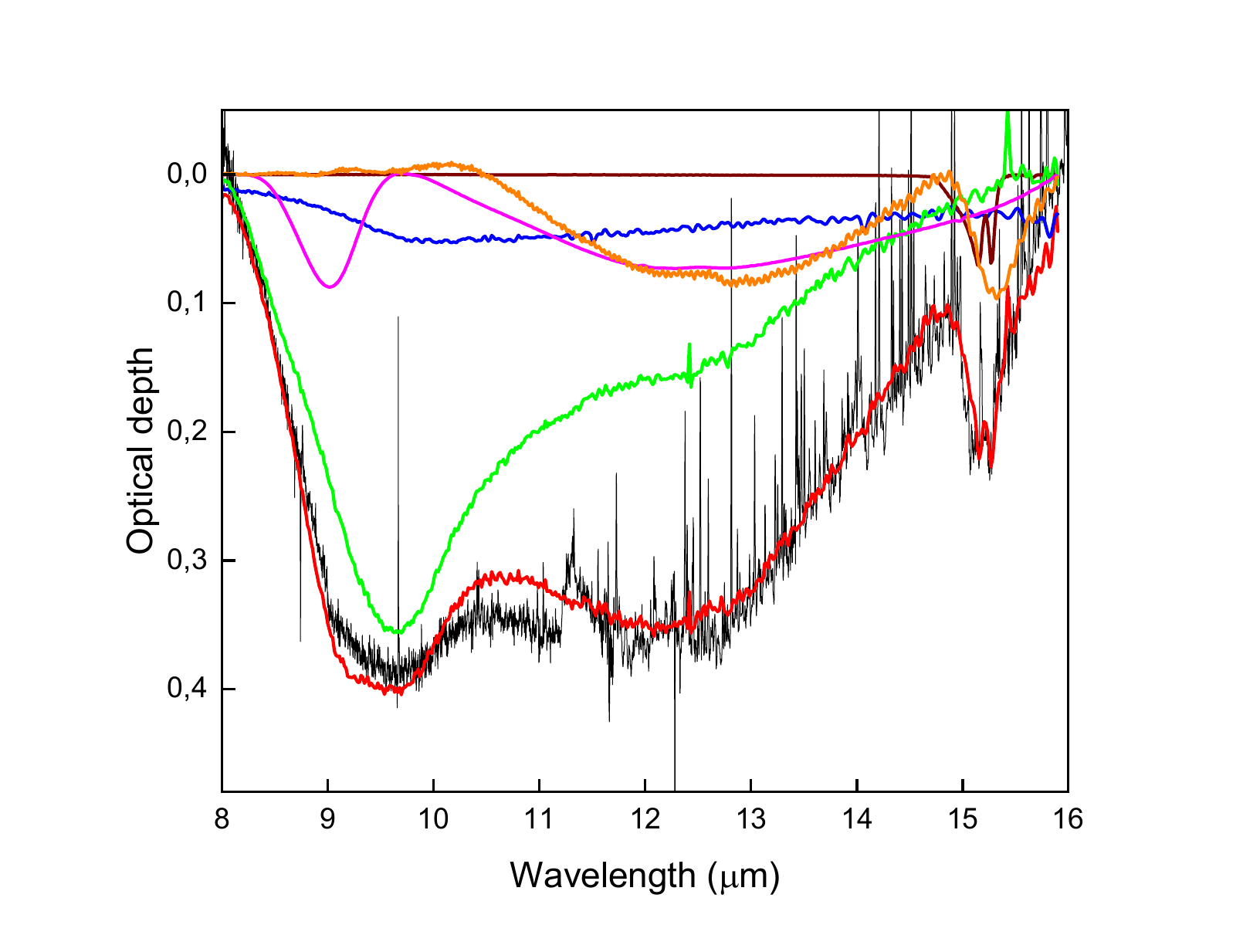}
      \caption{Observations (black), the best fit model for the 8 -- 16 $\mu$m region (red) and its components: MgSiO$_3$/H$_2$O 150 K (green),
Mg$_2$SiO$_4$/H$_2$O (blue),
H$_2$O/NH$_3$ (magenta),
H$_2$O/ CO$_2$ (orange),  CO$_2$ (brown).
              }
         \label{Fig:8.0-16.0}
   \end{figure}

This spectral region contains the silicate stretching band around 10 $\mu$m, the water librational band around 12 $\mu$m, and the  CO$_2$ bending band around 15 $\mu$m. The best fit model is presented in Fig.~\ref{Fig:8.0-16.0} and was obtained using the input mixture of the spectra of MgSiO$_3$/H$_2$O 150 K, Mg$_2$SiO$_4$/H$_2$O 150 K \citep[both spectra presented in][]{Potapov2021b},  CO$_2$, H$_2$O/ CO$_2$, and H$_2$O/NH$_3$ \citep[the spectra were presented in][]{Rocha2014,Rocha2017,McClure2023}, correspondingly.

We started with three-component models including  CO$_2$ and H$_2$O/ CO$_2$ spectra (basing on the  CO$_2$ results presented above) and 10, 100 or 150 K MgSiO$_3$/H$_2$O spectra. Inclusion of the second  CO$_2$ component was necessary to fit the  CO$_2$ band at 15 microns. We clearly observe the splitting of the band caused the segregation and/or distillation causing  CO$_2$ to be present in pure form as well \citep[cf.][]{Isokoski2013}. The MgSiO$_3$/H$_2$O 150 K spectrum provided the best fit from the three choices, and we used it in the following. At the next step, we added spectra of other silicates mixed with water ice, namely the
Mg$_2$SiO$_4$/H$_2$O 150 K and Mg$_2$SiO$_4$/H$_2$O 150 K spectra. Addition of the Mg$_2$SiO$_4$/H$_2$O spectrum led to an improvement of the fit. At the next step, we tried to add low temperature components, such as MgSiO$_3$/H$_2$O 10 and 100 K, H$_2$O/CH$_3$OH, H$_2$O/NH$_3$ and H$_2$O/CO/CO$_2$/NH$_3$/CH$_3$OH at 10 K, and high-temperature components, such as H$_2$O/CO$_2$/NH$_3$ at 150 K and H$_2$O/CO$_2$/NH$_3$/CH$_3$OH at
100 K. Only the H$_2$O/NH$_3$ addition improved the fit. Thus, a small amount of cold ammonia may be present in the disk ices. The AIC and RMSE values are presented in Table~\ref{table:6}.

\begin{table*}
\caption{EMIIGMA fitting results for the 8 -- 16 $\mu$m wavelength region.}             
\label{table:6}      
\begin{tabular}{l c c c c c c c c c c c c}
\hline\hline     
 & \multicolumn{12}{c}{Combination $\#$}  \\ 
 Component and Fit                    & 1 & 2 & 3 & 4 & 5 & 6 & 7 & 8 & \textbf{9} & 10 & 11 & 12  \\
 \hline \\
MgSiO$_3$/H$_2$O 10 K                 & x &   &   &   &   & x &   &   &            &    &    &    \\[1mm]
H$_2$O/CO$_2$                         & x & x & x & x & x & x & x & x & \textbf{x} & x  & x  & x  \\[1mm]
pure CO$_2$                           & x & x & x & x & x & x & x & x & \textbf{x} & x  & x  & x  \\[1mm]
MgSiO$_3$/H$_2$O 100 K                &   & x &   &   &   &   & x &   &            &    &    &    \\[1mm]
MgSiO$_3$/H$_2$O 150 K                &   &   & x & x & x & x & x & x & \textbf{x} & x  & x  & x  \\[1mm]
Mg$_2$SiO$_4$/H$_2$O 150 K            &   &   &   & x &   & x & x & x & \textbf{x} & x  & x  & x  \\[1mm]
MgFeSiO$_4$/H$_2$O 150 K              &   &   &   &   & x &   &   &   &            &    &    &    \\[1mm]
H$_2$O/CH$_3$OH$^a$                   &   &   &   &   &   &   &   & x &            &    &    &    \\[1mm]
H$_2$O/NH$_3$$^a$                     &   &   &   &   &   &   &   &   & \textbf{x} &    &    &    \\[1mm]
H$_2$O/CO/CO$_2$/NH$_3$/CH$_3$OH$^a$  &   &   &   &   &   &   &   &   &            & x  &    &    \\[1mm]
H$_2$O/CO$_2$/NH$_3$$^b$              &   &   &   &   &   &   &   &   &            &    & x  &    \\[1mm]
H$_2$O/CO$_2$/NH$_3$/CH$_3$OH$^c$     &   &   &   &   &   &   &   &   &            &    &    & x  \\[1mm]
AIC & 18.6& 16.2 & 14.2 & 15.5 & 16.2 & 17.5 & 17.4 & 17.4 & \textbf{15.6} & 17.9 &  17.9  & 17.5 \\[1mm]
RMSE & 0.050 & 0.045 & 0.040 & 0.038 & 0.040 & 0.038 & 0.038 & 0.038 & \textbf{0.033} & 0.039  & 0.039 & 0.038 \\
\hline  \\                
\end{tabular}
\tablefoot{The Akaike Information Criterion (AIC) and Root-mean-square error (RMSE) for various input mixtures are listed. The best fit parameters are marked in boldface. \\
\noindent
\tablefoottext{a}{ice mixture at  10 K} \,\,
\tablefoottext{b}{ice mixture at 150 K} \,\,
\tablefoottext{c}{ice mixture at 100 K}
}
\end{table*}

\section{Discussion}

\subsection{NH$_3$ chemistry and complex organic molecules}

Ammonium carbamate is a product of the NH$_3$ +  CO$_2$ chemistry (reaction  CO$_2$ + 2NH$_3$ $\rightarrow$ NH$_4^+$NH$_2$COO$^-$), which, as it was shown by laboratory experiments, can be triggered by heat \citep[above 60 -- 65 K,][]{Bossa2008,Potapov2019,Potapov2020} and much less efficiently by energetic processing \citep{Bossa2008}. The detection of ammonium carbamate is consistent with the detection of NH$_4^+$ and OCN$^-$, also the products of NH$_3$ chemistry, where NH$_3$ reacts with HNCO leading to the production of both species via a direct pure thermal acid-base reaction or with CO leading, if energetically triggered (UV photons, ions), to the production of OCN$^-$ \citep{Novozamsky2001,Martinez2014}.
NH$_4^+$ can be also produced by ultraviolet photolysis of NH$_3$--H$_2$O ice mixtures \citep{Moon2010} and, in principle, can be a product of dissociation of ammonium carbamate.

Pure thermal reactions definitely take place in planet-forming disks, thus, the formation of all three species (NH$_4^+$NH$_2$COO$^-$, NH$_4^+$, OCN$^-$) can be directly related to the disk chemistry. The detection of these molecules is consistent with a questionable detection of NH$_3$ and non-detection of warm  CO$_2$, clearly showing their efficient involvement into the disk chemistry and their evolution tracks. Energetic processing may take place in both, the disk and its parent molecular cloud environments, however, midplanes and warmer layers of planet-forming disks -- the locations of molecular ices, are shielded from external UV fields. An open question is the survival of the molecules, if formed in the parent molecular cloud, during the chemical evolution of the environment (transformation of a cloud into a disk). Dissociation, desorption and further reactions of the species may lead to their disappearance. Moreover, ammonium carbamate has not been detected in molecular clouds. Thus, the synthesis of the detected species can be referred to both the disk studied and the molecular cloud from which it was formed with a clear preference to the disk.

One of the fundamental questions regarding small Solar System bodies (comets, asteroids and their meteoritic remains) is the origin of detected complex organics, particularly building blocks of biological macromolecules, such as sugars, amino acids, and nucleobases. Whether these species originate from the parent to the Solar System protoplanetary disk or even the interstellar cloud is a long-standing discussion in the astrochemical community. Many complex organic molecules have been detected in comets. Our particular attention attracted the detection of ammonium salts, such as ammonium carbamate (NH$_4^+$NH$_2$COO$^-$), ammonium formate
(HCOO$^-$NH$_4^+$), and ammonium sulphate (NH$_4^+$)$_2$SO$_4^{2-}$ \citep{Poch2020}. Ammonium salts may dominate the reservoir of nitrogen in comets and asteroids and, considering previous exogenous delivery hypotheses \citep[e.g., ][]{Oro1961,Brack1991}, on the early Earth. Being delivered to the early Earth and mixed in liquid water, ammonium salts could participate in prebiotic reactions leading to the synthesis of amino acids, nucleobases and sugars \citep{Lerner2005,Dziedzic2009,Callahan2011}. For instance, ammonium carbamate can be a precursor of urea CO(NH$_2$)$_2$ and urea, in turn, is a possible precursor of pyrimidine C$_4$H$_4$N$_2$ required for the synthesis of nucleobases in RNA molecules \citep{Robertson1995}. Thus, the detection of ammonium carbamate in the protoplanetary disk links parent protoplanetary and daughter planetary environments and brings us one more step closer to the understanding of the processes and conditions leading to the formation of prebiotic worlds. Moreover, as the Solar System is believed to be formed in a massive star-forming region, the Orion disks most likely represent analogues of the protoplanetary disk phase of the Solar System \citep[e.g., ][and references therein]{Young2020,Dauphas2011}. The detection of ammonium salts in the Orion disks can then be directly linked to their detection in Solar System objects like the comet 67P.

\subsection{Warm vs Cold}

Highly inclined disks, which are not completely edge on (inclination angles of 70 -- 85 degrees) present a particular interest. As shown for the HH 48 NE disk \citep[see][]{Sturm2024}, the location of the ices detected in the NIRSpec range is mainly warm layers significantly above the midplane. This is consistent with our detections of warm water ice (100/150 K) at 3 microns, warm NH$_4^+$ (80 K) at 7 microns, and ammonium carbamate (a result of ``warm'' chemistry). Detection of cold CO speaks for its presence only in the midplane of the disk due to its very low desorption temperature (around 30 K) and \citet{Sturm2024} had the same view. Detection of only cold  CO$_2$ (having the desorption temperature of 85 K) points to its presence only near the midplane of the disk and its depletion in the warmer layer of the disk due to the efficient NH$_3$ +  CO$_2$ chemistry discussed above. In the 10-micron range, the absorption is predominantly from regions near the disk midplane \citep[cf.][]{Sturm2024}, and here we detect a mixture cold and warm materials.

\section{Conclusions}

The highly inclined disk d216-0939 has been observed by JWST. The analysis of the absorption bands shows the presence of simple and complex molecular species. For the first time, a complex organic molecule ammonium carbamate has been detected in astrophysical environments pointing to the complex chemistry taking place in
protoplanetary disks. Together with the detection of ammonium salts in the Solar System Comet 67P, this result links protoplanetary disks and minor bodies of the Solar System.

\begin{acknowledgements}
We thank the referee for a constructive report. We are grateful to the JWST team for the wonderful observations. This work is based on observations made with the NASA/ESA/CSA James Webb Space Telescope. The data were obtained from the Mikulski Archive for Space Telescopes at the Space Telescope Science Institute, which is operated by the Association of Universities for Research in Astronomy, Inc., under NASA contract NAS 5-03127 for JWST. These observations are associated with program \#1741. The study was supported by the Federal Ministry for Economic Affairs and Climate Action on the basis of a decision by the German Bundestag (the German Aerospace Center project 50OR2215). A.~P. acknowledges support from the Deutsche Forschungsgemeinschaft (Heisenberg grant PO 1542/7-1).
\end{acknowledgements}

\bibliographystyle{aa}
\bibliography{water.bib}

\end{document}